\documentclass[]{spie}  

\usepackage{amsmath,amsfonts,amssymb}
\usepackage{graphicx}
\usepackage[colorlinks=true, allcolors=blue]{hyperref}
\usepackage{wrapfig}

\title{STARI: STarlight Acquisition and Reflection toward Interferometry}

\author[a]{John D. Monnier}
\author[b]{Prachet Jain}
\author[b]{Shashank Kalluri}
\author[b]{James Cutler}
\author[c]{Simone D'Amico}
\author[d]{Glenn Lightsey}
\author[e]{Leonid Pogorelyuk}
\author[f]{Gautam Vasisht}
\author[g]{Kerri Cahoy}
\author[a]{Michael Meyer}
\affil[a]{Astronomy Department, University of Michigan, Ann Arbor, MI, USA}
\affil[b]{Aerospace Engineering Department, University of Michigan, Ann Arbor, MI, USA}
\affil[c]{Stanford University, Palo Alto, CA, USA}
\affil[d]{The Georgia Institute of Technology, Atlanta, GA, USA}
\affil[e]{Rensselaer Polytechnic Institute, Troy, NY, USA}
\affil[f]{Jet Propulsion Laboratory, Pasadena, CA, USA}
\affil[g]{Massachusetts Institute of Technology, Cambridge, MA, USA}

\authorinfo{Further author information: (Send correspondence to J.D.M., E-mail: monnier@umich.edu)}

\begin{document} 
\maketitle

\begin{abstract}
We present the concept for STARI: STarlight Acquisition and Reflection toward Interferometry. If launched, STARI will be the first mission to control a 3-D CubeSat formation to the few mm-level, reflect starlight over 10s to 100s of meters from one spacecraft to another, control tip-tilt with sub-arcsecond stability, and validate end-to-end performance by injecting light into a single-mode fiber.  While STARI is not an interferometer,  the mission will advance the Technology Readiness Levels of the essential subsystems needed for a space interferometer in the near future.  
\end{abstract}


\section{Introduction}
\label{sec:intro}  

The angular resolution of a telescope is fundamentally limited by the diffraction of light, imposing a frustrating limit to what we can learn about the Universe.  On the ground, we have overcome these strictures through the technique of {\em interferometry}, whereby light beams from a far-flung array of telescopes are combined to achieve the imaging resolution of a giant mirror the size of the array itself \cite{monnierreview2003}. In recent years, the CHARA interferometer has imaged the surfaces of stars beyond the Sun with {\em milli-arcsecond} angular resolution \cite{roettenbacher2016} while the Event Horizon Telescope has combined mm-wave light captured across the globe to dissect the event horizon of a black hole at {\em micro-arcsecond scales} \cite{eht2019}.  

While geography, weather, and atmospheric turbulence all limit the reach of ground-based interferometers, the scientific potential of {\em space} interferometry is truly vast.  In space, there are no physical limits to the telescope spacings, thus any angular resolution is theoretically accessible. Further, without the blurring of turbulence, long coherent integrations and high sensitivity are also achievable.  {\bf A mid-infrared space interferometer could null out light from nearby stars and detect biomarkers in the atmospheres of Earth-like planets} \cite{tpfi2008} while an X-ray interferometer could image the accretion disks around black holes and neutron stars \cite{maxim2004}.  NASA and ESA are actively pursuing related technologies to enable a space-based gravitational wave observatory called LISA, the Laser Interferometer Space Antenna Mission, with three spacecrafts located millions of kilometers apart flying in a precision formation.

Over the past 20 years, most of the sub-systems needed for space interferometry have been developed and tested, such as propulsion \cite{arestie2012}, laser metrology \cite{gracefo2019}, space GPS,  and nulling beam combiners \cite{kg2017}.   However, no system-level demonstration of formation flying has been attempted at the level needed to coherently combine light, largely due to the high costs and risk of even a basic mission.  Fortunately, this situation has radically changed with the revolution in space accessibility made possible by SmallSats and CubeSats. 

Here, we describe a two-CubeSat pathfinder called STarlight Acquisition and Reflection toward Interferometry (STARI)  to advance critical {\em system-level} technologies for space interferometry built upon the mature CubeSat platform.    STARI will fill the few remaining gaps in our technology readiness to permit a science-based SmallSat interferometer mission to closely follow.  The 6U CubeSat STARI-1 will host a small 2.5cm siderostat with precision guiding to redirect starlight across hundreds of meters to the STARI-2 spacecraft, an identical partner to STARI-1 which will then also send light to STARI-1.   Each STARI will also launch a collimated laser reference to guide starlight into a single-mode fiber, a critical building block for a future nulling interferometry combiner needed for exoplanet science. An advanced sensor fusion approach will combine differential GPS and multi-LED beacon imaging to measure relative positioning of STARI-1 and STARI-2 to within a few millimeters.  Mission Operations will also validate propulsion and thruster performance for orientation reconfiguration, following passively-safe orbital evolution.   STARI relies on high heritage and we believe can be built with acceptable risk.

In this report, we outline the motivating science goals for developing space interferometry, describe the current formation-flying space context, and outline the basic features of STARI.

\section{Key Science Driver: Earth-like planets in the mid-infrared}
To motivate our project, we outline how a mid-infrared long-baseline interferometer is perhaps our best hope to  discover evidence for life around nearby terrestrial Earth-like planets.

\begin{figure}[!b]
\centering
\vspace{-.2in}
\includegraphics[width=5.75in]{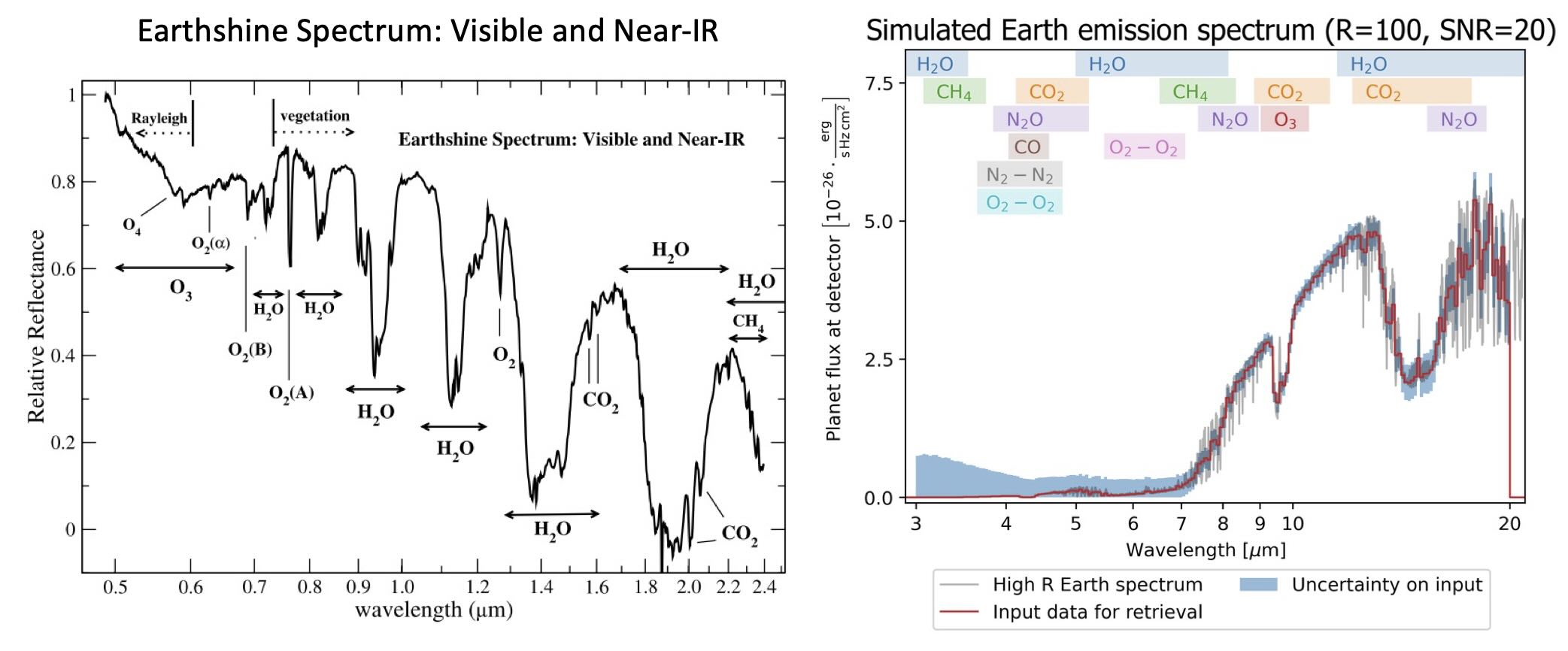}
\vspace{-.15in}
\caption{Characterizing atmospheres of planets requires both visible and mid-IR spectra  (left). The moon's Earthshine was measured \cite{turnbull2006} and showed molecular species in lunar-reflected light of an ``Earth'' (right).  Earth-like exoplanets emit the bulk of their energy in the mid-IR, a part of the spectrum filled with signatures of molecules such as water, methane, ozone, and other potential ``biomarkers.'' (image credit: life-space-mission.com). 
  \label{fig_lifeb}}
\end{figure}

Let's consider the next ten years of exoplanet science. We will learn about the outer planetary system architectures from Gaia astrometry and the Roman Space Telescope microlensing surveys.  Our initial sample of habitable Earth-like planets will likely come from transit surveys,  but these systems will be typically about $6\times$ farther away than the closest comparable objects due to orbital alignment selection effects making followup difficult. Long duration radial velocity surveys of nearby stars will be completed and perhaps will turn up a handful of Earth analogues, but only around the most stable  stars. 
Thus, despite decades of effort, {\it our knowledge of Earth analogues around Sun--like stars within 10 pc will remain extremely limited} for the foreseeable future. Even ambitious space-based imaging missions slated for launch $>$2035 (such as Habitable Worlds Observatory HWO) might (best case) only detect $\sim$10 rocky planets in the liquid-water zone of Sun--like stars.

Long--baseline mid--infrared interferometry offers an extremely attractive path that is complementary to the visible coronagraphic missions like HWO, one that leads to a whole that is greater than the sum of the parts (see Figure~\ref{fig_lifeb}).  Indeed, studies\cite{alei2022} show that visible or mid-infrared observations alone are each blind to important atmospheric constituents and a broad wavelength coverage is required to understand the chemical roles of potential biomarkers.  Thermal emission spectra also provides the key to unlock the radius--albedo ambiguity for cloudy planets that are only observed in the visible, and thus an assessment of the global energy budget of an Earth--like planet. Only a joint analysis of visible and infrared data can decisively constrain a self--consistent model of planetary climate, recovering temperature--pressure profiles, including an understanding of surface temperatures and assessment of global ocean coverage.  These are the necessary ingredients to accurately assess habitability and to interpret potential biosignatures, not to mention to fully characterize the great diversity of planetary atmospheres we will encounter, based on heritage from JWST and ARIEL.

\begin{wrapfigure}[18]{r}{3in}
\centering
\vspace{-.15in}
\includegraphics[width=3in]{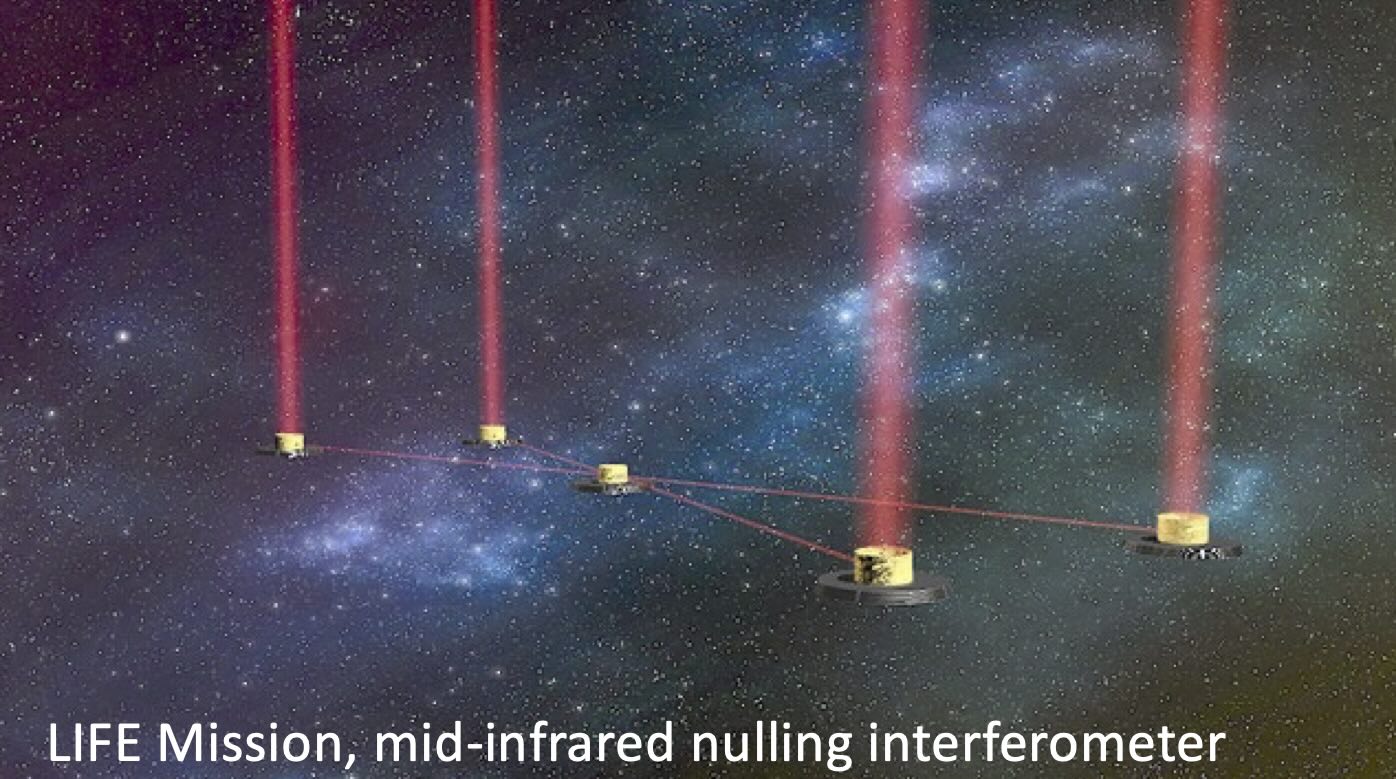}
\vspace{-.05in}
\caption{ The Large Interferometer For Exoplanets (LIFE)\cite{2021LIFEI} is a proposed mission being considered  by the ESA Voyage 2050 program. The design for an X-shaped nulling interferometer has direct legacy from NASA Terrestrial Planet Finder Interferometer (TPFI) studies, though other architectures are under consideration.
\label{fig_life_concept}}
\end{wrapfigure}

While NASA is developing  a challenging
coronagraphic imaging mission in the visible (HWO),  the European Space Agency has started planning for a potential mid-infrared free-flying interferometer.  Envisioning missions for the 2035-2050 time frame, the ESA Voyage 2050 program is considering the concept behind the Large Interferometer for Exoplanets (LIFE) mission\cite{2021LIFEI} (see Figure~\ref{fig_life_concept}).

While the focus here is on exoplanets, a mid--infrared space interfermeter would also be able to make fundamental discoveries in other areas.  Dissecting planet forming disks as a function of host star properties will help us understand how circumstellar disk chemistry and dynamics intertwine to produce the diversity of observed planetary systems, a topic studied recently by the {\em Planet Formation Imager} (PFI)\cite{pfi2018}.  Watching dust form in the outer layers of massive stars and evolved stars will help us understand how carbon and other volatile species are recycled from stars to the interstellar medium and into forming planets.  Zooming into the circum--nuclear environments of active galaxies, disentangling star formation from accreting black holes, and the co--evolution of galaxies are all possible use cases.  Even multi--wavelength imaging of strong gravitational lensing of distant quasars can reveal sub--structure in the matter distribution of lensing galaxies, and/or explore variability along different sight lines as a geometric probe of Hubble expansion.

It is said ``Space interferometry is inevitable'' and new technologies make space interferometry more feasible than ever. Investment now will save in cost, risk, and schedule for the inevitable pathfinder missions to come. We will now describe the STARI Mission, a modest first step down this exciting path.

\section{Current Context:  SmallSat and CubeSat Platforms}
\begin{figure}[!t]
\centering
\includegraphics[width=6.25in]{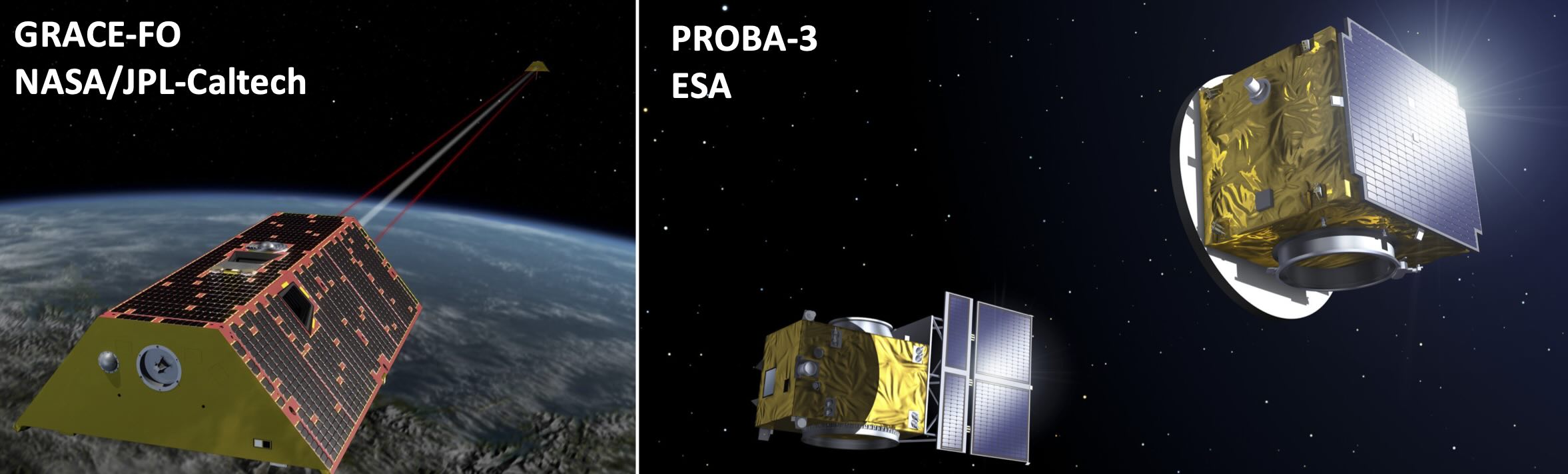}
\caption{The GRACE-FO Mission (gracefo.jpl.nasa.gov) has demonstrated precision metrology and the upcoming PROBA-3 (www.esa.int) will further mature formation flying technologies. Similar capabilities are now being flown using smaller CubeSats. 
  }\label{fig_SmallSats}
  \vspace{-.1in}
\end{figure}

While we may still be a decade from first fringes with a formation-flying space interferometer, tangible advances in crucial technologies are taking place now, made possible by the radically-reduced costs of developing and launching small satellites.  A recent white paper\cite{monnier2019_whitepaper_ff}  outlined some of the formation flying technologies that have already been demonstrated, most notably by the GRACE-FO mission that used a long coherence length laser to measure the separation between two spacecraft in low earth orbit and the much-anticipated upcoming PROBA-3 mission (see Figure~\ref{fig_SmallSats}; launch 2024).

\begin{wrapfigure}[26]{L}{3.5in}
\centering
\vspace{-.1in}
\includegraphics[width=3.5in]{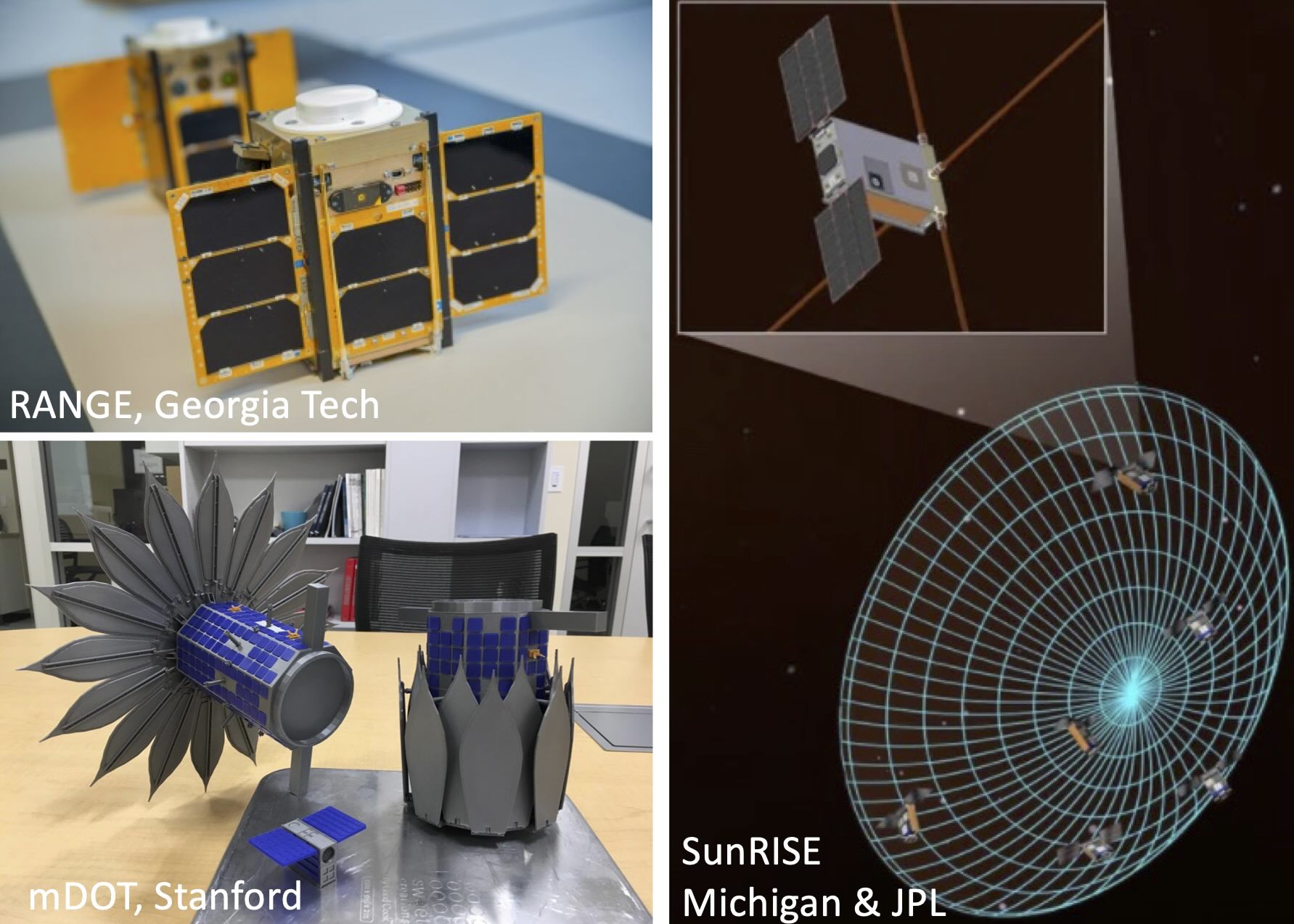}
\vspace{-.1in}
\caption{This figure highlights just a few of the formation-flying CubeSat missions in development that inform the design of STARI.  (top-left) Some RANGE missions (PI: Gunter, Georgia Tech) have already flown and  consisted of two CubeSats testing multiple ranging techniques and novel propulsion.  (bottom-left) The mDOT mission is under study to fly a starshade pathfinder using formation-flying CubeSats (PI: Macintosh, Stanford/UCSC).  (right) The SunRISE mission is a fully-funded formation-flying radio long-wave radio interferometer using 6 CubeSats (PI: Kasper, U. Michigan + JPL).
\label{fig_CubeSats}}
\vspace{-.2in}
\end{wrapfigure}

GRACE-FO and PROBA-3 both use SmallSats ($\sim$100-500kg), a significant step above the CubeSat ($\sim$10kg) architecture we are considering for STARI.  Some power hungry and large-volume technologies can not currently be deployed in the much-smaller CubeSat platform where a 1U unit is only 10cm x 10cm x 10cm and where 6U to 12U are typical sizes.  That said,  increasingly ambitious CubeSat missions are being proposed and flown, including to test laser metrology, precision star-tracking,  fine propulsion control, and even formation-flying of up to 6 CubeSats (see Figure~\ref{fig_CubeSats}).

Certainly our group is not alone in seeing the immense promise of CubeSats for formation flying (see Figure~\ref{fig_CubeSats}).  According to a recent paper\cite{bandy2016}, fourteen multi-spacecraft nanosat ($<$10kg) missions have qualified as ``formation flying'' and an additional 25 have  flown in a looser ``constellation'' configuration.  In 2018, NASA Astrophysics chose two CubeSat formation flying missions for additional study, mDOT (starshade) and VTXO (X-ray imaging). Launched in July 2023, the NASA Starling is successfully demonstrating ``Autonomous Swarm Technologies'' using CubeSats.  Lastly, we highlight the SunRISE Mission, because it will be the first mission to attempt CubeSat interferometry, with an expected 2024 launch.

The NSF-funded VISORS Mission is a CubeSat formation flyer for extreme-UV solar imaging on-track to be launched in 2024.  VISORS\cite{koenig2023} has developed advanced thrusting and differential GPS technology to permit a two 6U CubeSats to fly 40m apart with cm-level relative positioning.  {\em Indeed, one critical argument we put forward to justify the technology readiness for STARI is the  launching of a flood of SmallSat formation flying missions currently under development: Starling, RACE, SWARM-EX, Aero-Vista, CLICK, VISORS, SunRISE, and PROBA-3.}  While many of these missions have not flown yet, we expect that many will have in time to inform the final design of STARI. 

While a CubeSat optical interferometer would be a quantum leap in capability, it only requires a modest improvement in technology over missions described above and those demonstrated on the ground \cite{ireland2020}.  In the 2000s, NASA-JPL developed the ``Formation Control Testbed'' (see \url{http://dst.jpl.nasa.gov/test\_beds/}), with 6-dof robots moving on a plane with freely-rotating ``spacecraft.''  They developed advanced simulation software validated with a real-world lab demonstration. More recently, M. Ireland (ANU) has started building the  Pyxis testbed (\url{http://www.mso.anu.edu.au/pyxis/}) that uses portable and moving robots to simulate 6-dof motions of CubeSat telescopes reflecting starlight to an immobile central station for true star tracking and interferometric beam combination.  Similarly, J. Monnier's group at Michigan is using drones to test long-baseline formation flying concepts (see contribution in these proceedings).  There has also been subsystem level description of low-power metrology systems \cite{lagadec2020} and novel nulling combiners suitable for space \cite{martinod2021}. Lastly, we note other independent ideas for feasible smallSat interferometers \cite{hansen2020,dandumont2021,matsuo2022} in low-earth orbit.  
 
 The STARI mission will build upon and complement these pioneering efforts. 
 
\vspace{-.1in}
\section{The STARI Mission}

Having established the strong science potential for space interferometry and the significant progress in the required technologies over the past decade, here we develop the proposed STARI Mission.
STARI, STarlight Acquisition and Reflection toward Interferometry, will be the first mission to control a 3-D CubeSat formation to the few mm-level, reflect starlight over 10s to 100s of meter from one spacecraft to another, control wavefront tip-tilt with sub-arcsecond stability, and validate end-to-end performance by injecting light into a single-mode fiber.  These four major steps are essential to any future space interferometer and by increasing their Technology Readiness Levels (TRLs) we will pave the way to a science mission as a next step.  STARI has been designed to be identical to one arm of an interferometer, which would leave only the space-qualified nulling interferometric combiner payload and space-borne delay line subsystems at low TRL. 
\label{stari}

\begin{figure}[!t]
\centering
\caption{STARI top level requirements for baseline and extended mission.
  }\label{fig_tlrs}
\includegraphics[width=6.25in]{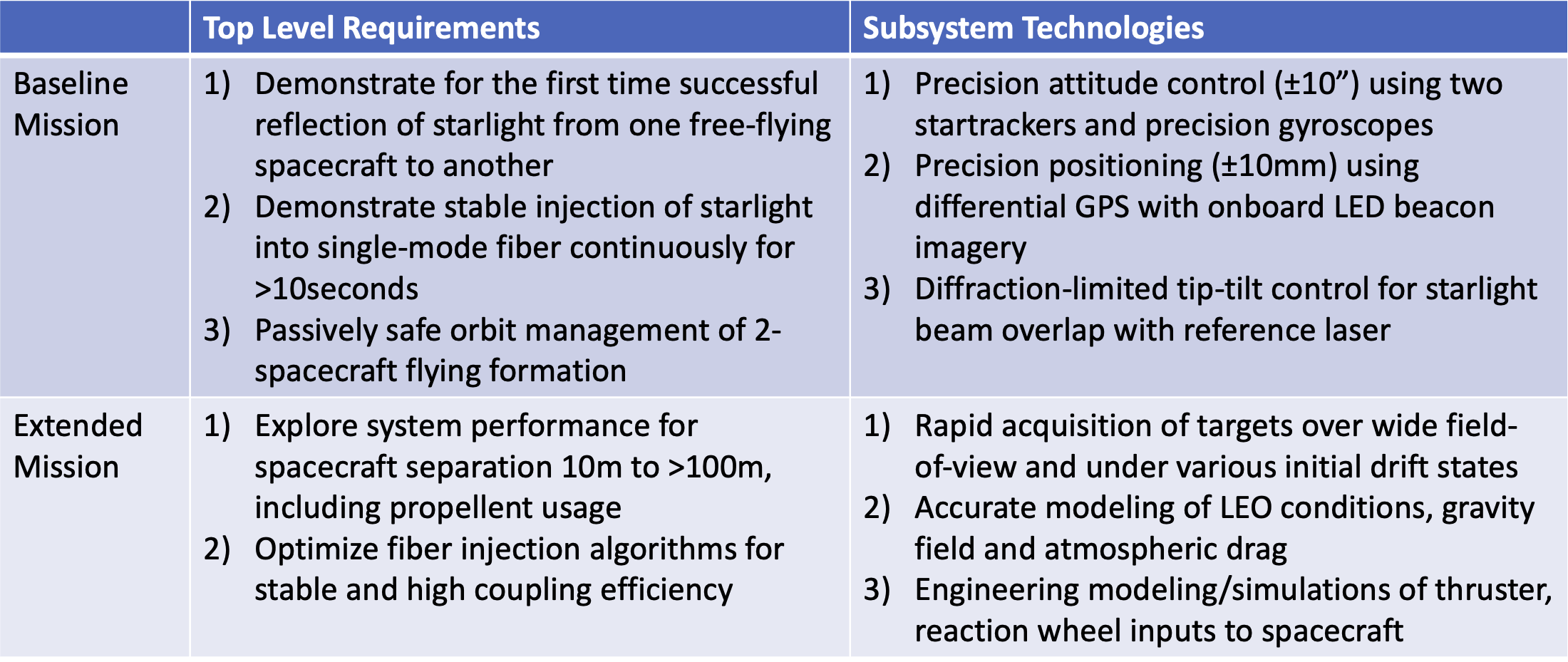}

  \vspace{-.1in}
\end{figure}

\subsection{Mission Goals}
\begin{wrapfigure}[21]{r}{4.7in}
\vspace{-.55in}
\centering
\includegraphics[width=4.5in]{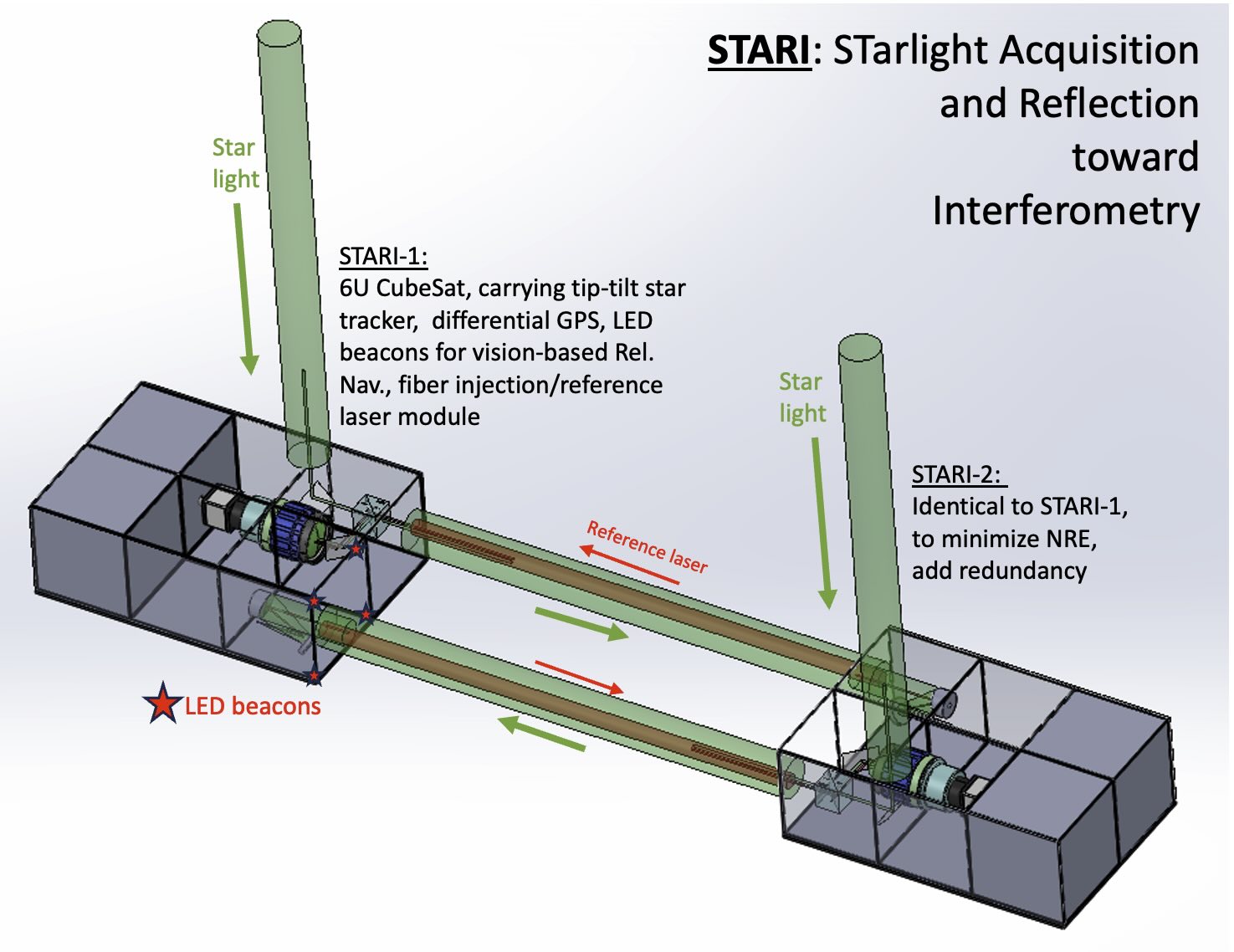}
\vspace{-.1in}
\caption{Model of STARI-1 and STARI-2 CubeSats flying in formation. We show the beam paths and sketch the footprints of the primary payloads }\label{fig_overview}
\end{wrapfigure} 
We have translated these goals into STARI Top level Requirements (TLRs) with associated subsystem requirements.  Figure~\ref{fig_tlrs} is split into the TLRs for the Baseline Mission and for an Extended Mission.  Achieving the Baseline Mission would demonstrate the feasibility for a formation flying space interferometer while the Extended Mission would validate engineering flight and operations modeling to further retire risk for future science interferometers.  In order to transfer a beam of starlight from one spacecraft into a fiber optic located on a another spacecraft, one has to have exquisite control of {\em both attitude and position} of the two formation-flying spacecraft.  We now will outline our concept of operations, the flowdown to technical requirements, and consider the risks.

\subsection{STARI Architecture}

A schematic of the mission concept is shown in Figure~\ref{fig_overview} to guide our discussion. To orient readers, let us follow light from the star to the detector in our preliminary design reference mission (DRM).  The STARI-1 6U CubeSat collects light with 2.5cm siderostat mirror with precision (sub-arcsecond) tip-tilt actuators.   A differential GPS system with precision gyroscopes will deliver few-cm relative positioning and sub-arcinute orienteering, while a 2nd startracker and vision-based LED beacon system will further improve our formation knowledge to few-mm relative positions and few-arcseconds in roll, yaw, and tilt.  At this point, STARI-1 will receive a boresight laser reference from STARI-2 to guide starlight into the STARI-2 off-axis parabola (OAP) aperture.   These data along with traditional accelerometers and gyroscopes will be combined via sensor fusion algorithms to direct reaction wheels and thrusters (based on design from VISORS) to maintain the formation.  By overlapping starlight and a laser reference on a  beamsplitter, we can reliably inject the starlight into a single-mode fiber using the fine tip-tilt system on the siderostat.  The starlight will be detected with a fiber-coupled,  single-pixel photodiode to allow measurement of the coupling stability, a critical input for mission modeling of a future science-capable interferometer. 

Let's discuss each subsystem below:

\begin{wraptable}{r}{3in}
\vspace{-.2in}
\caption{Overview of STARI Parameters \label{table_params}\\}
\vspace{.05in}
\begin{tabular}{|p{1in}|p{1.5in}|}\hline
 Parameter   & Value \\
 \hline
 Volume & 2x6U \\
 Mass & 2x10kg (nominal) \\
 Power & 20W (nominal) \\
 Orbit & LEO, $>$500km \\
 Lifetime & 6 months \\
 Separations  & 10-100m \\
   \hline
\end{tabular}
\end{wraptable}
\noindent{\bf Spacecraft Bus:} We have already contacted commercial vendors Blue Canyon Technologies, Terran Orbital, and Endurosat to confirm availability of relevant spacecraft busses for STARI in an appropriate price range. Given the similarity to formation-flying VISORS mission, we have confidence our bus requirements will be modest (see Table\,\ref{table_params}), including conventional solar panels, integrated power distribution, ground communication, and Attitude Determination and Control System (ADCS). To simplify interfaces, add redundancy, and reduce non-recurring engineering expenses with spacecraft vendors, we anticipate that STARI-1 and STARI-2 spacecraft will be identical.

\noindent{\bf Relative Navigation (GPS):} Coarse position knowledge (meter-level) is needed for spacecraft safety and fine knowledge (cm-level) is required to successful reflect light from one spacecraft to another.  The Stanford team led by Simone D'Amico has developed a sophisticated GNC system that uses the low-level telemetry from commercial GPS units to allow cm-level relative knowledge in low Earth orbit that is core to a number of on-going missions, most notably VISORS (launch 2024).  Of specific relevance here has been the development of the Distributed Multi-GNSS Timing and Localization (DiGiTaL)\cite{giralo2021} coupled with the development of the closed-form impulsive control algorithms for formations such as TANDEM-X and PRISMA \cite{gaias2015}.

\begin{wrapfigure}[26]{r}{4.0in}
\vspace{-.2in}
\centering
\includegraphics[width=3.9in]{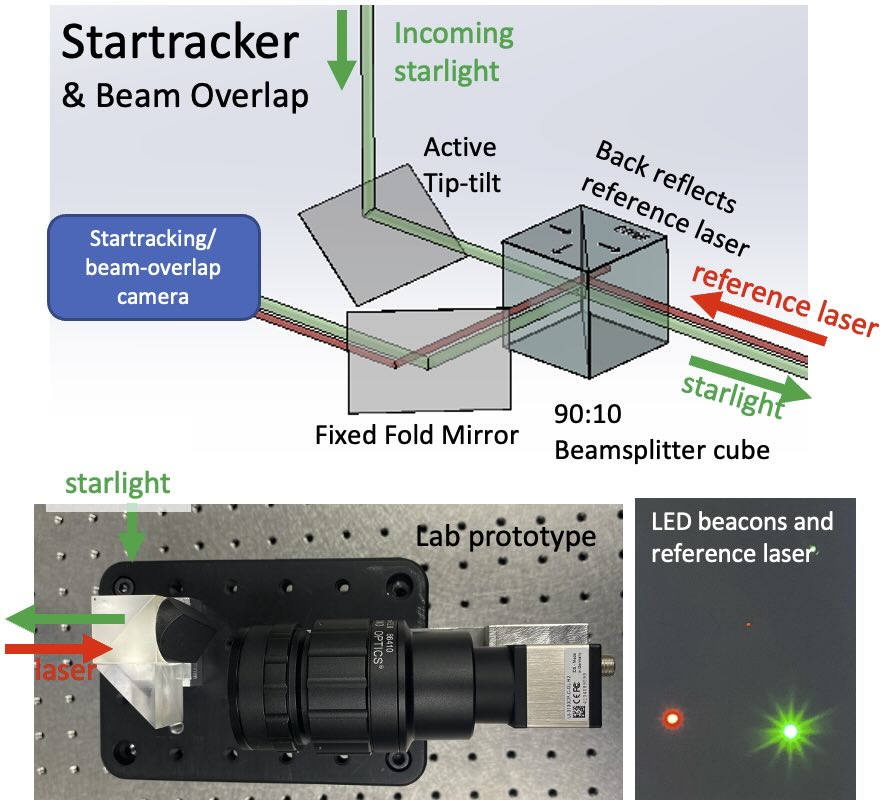}
\vspace{-.0in}
\caption{Optical design (top) and lab prototype (bottom-left) of the STARI startracker. The camera frame shows the complex image containing LED beacons (for mm-level positioning) and a laser reference spot used to guide starlight into the single-mode fiber (see Figure~\ref{fig_stari2}).  }\label{fig_startracker}.
 \vspace{-.0in}
\end{wrapfigure}

\begin{wraptable}[8]{r}{3.6in}
\vspace{-.1in}
\caption{STARI Formation Stability Goals}
\begin{tabular}{|p{2.75in}|p{0.5in}|}\hline
 Parameter   & Value \\
 \hline
 GPS-based relative position (mm) & 15 \\
 LED-beacon relative position (mm) & 1 \\
 Relative Velocity (mm/s) & 0.1 \\
 Attitude Stability (arcsec/s) & 1 \\
 Impulse control (1 bit, mm/s) & 0.04 \\
  \hline
\end{tabular}
\end{wraptable}
\noindent{\bf Precision Attitude Sensing:} Positional  and attitude knowledge from GPS and conventional gyroscopes are not precise enough to accomplish STARI's primary goal of reflecting starlight from one spacecraft to another. For an interferometer, we need precision yaw, tilt and roll that will be derived from having two startrackers onboard each spacecraft.  The STARI-1 startracker will point toward our target star, which will allow precise (few arcsecond) tracking in yaw and tilt, however without additional sensors, the roll will not be well constrained beyond the gyroscope system.  Precision roll will be picked up using the spacecraft's ADCS startracker oriented roughly 90 degrees away.  We present our preliminary optical design of our startracker system in Figure~\ref{fig_startracker} including the beamsplitter hardware needed for the LED beacon and reference laser overlap functionality (explained below).

\noindent{\bf Precision Relative Navigation (LED beacons):} 

With arcsecond knowledge of spacecraft attitude, we can use a side-looking camera to view an LED constellation on each spacecraft to measure relative positions. Note that without precision attitude, we could not  distinguish between a relative attitude or positional offset.  Precision centroiding of the LED spots will allow mm-level precision for spacecraft separations out to 100 meters (1mm/100m $=$2 arcseconds).

In order to minimize the mass and power requirements, the same camera will view the star and LED beacons using a beamsplitter (see Figure \,\ref{fig_startracker}). This beamsplitter will also fix the angle of the STARI-1 and STARI-2 baseline to be nearly perpendicular to our line-of-sight, an important property for a future interferometer where we must match the optical path differences for starlight that is interfered at a central combiner.  
Figure~\ref{fig_startracker} shows a startracker lab prototype, where we have demonstrated centroiding with $<$1/20 pixel errors, corresponding to a positional uncertainty $<$1mm for our testing conditions using a 60-meter baseline. 
See Monnier et al. SPIE paper in these proceedings exploring how drones can be used as part of integration and testing for formation flying CubeSat missions.

\begin{wrapfigure}[17]{r}{3.65in}
\vspace{.05in}
\centering
\includegraphics[width=3.6in]{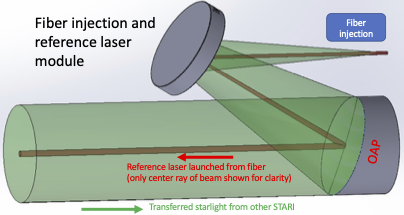}
\vspace{-.05in}
\caption{Optical design and prototype of the STARI fiber injection and reference laser modules (shown as pencil beam for clarity). Starlight is focused into a single-mode fiber while an outgoing laser reference is introduced using a fiber tap (not shown).   }\label{fig_stari2}.
\end{wrapfigure}

\noindent{\bf Single-mode Fiber Injection:         } \\           
While an interferometric beam combiner itself is not included in STARI, we do anticipate single-mode fibers will be used in a future interferometer mission due to their spatial filtering, stability,  and compatibility with integrated optics\cite{mirc2004,hsiao2010,martinod2021}.  Thus, a primary mission for STARI is stable injection of starlight into a single-mode fiber.  Figure~\ref{fig_stari2} shows our design that will use an off-axis-parabola (OAP) to focus the collimated starlight into a fiber.  We will send out a reference laser using back-illumination, injecting laser light using an unbalanced fused coupler where a small fraction of light is sent out the fiber and is collimated by the off-axis-parabola. As explained already, this reference laser is used in a tip-tilt loop to send starlight back along this path.  While some laser might backscatter into the starlight fiber, we can use a narrow laser optical filter to remove this contamination before detection.  

\noindent{\bf Starlight/Reference Laser Coalignment:} A main challenge for a future formation flying interferometer is the diffraction-limited  overlapping required for the telescope beams entering the interferometric combiner.  Here, we will solve this problem by launching a boresight laser reference from STARI-2,  co-aligned with the fiber injection optics.  The reference laser will be detected by the startracker camera using an internal reflection with a 90:10 beamsplitter on STARI-1.   With this optical arrangement (see Figure~\ref{fig_startracker}), we can enforce precise beam overlap of the starlight with the reference laser using a fine-level tip-tilt correction on the siderostat, guaranteeing the starlight will inject into the single-mode fiber.  This level of tip-tilt control with a CubeSat has been demonstrated by ASTERIA \cite{smith2018,kennedy2020} and multiple tip-tilt technologies exist (e.g., deformable mirrors)\cite{morgan2021}.

\noindent{\bf Propulsion:} 
\begin{wrapfigure}[24]{r}{3.0in}
\vspace{-.35in}
\centering
\includegraphics[width=2.9in]{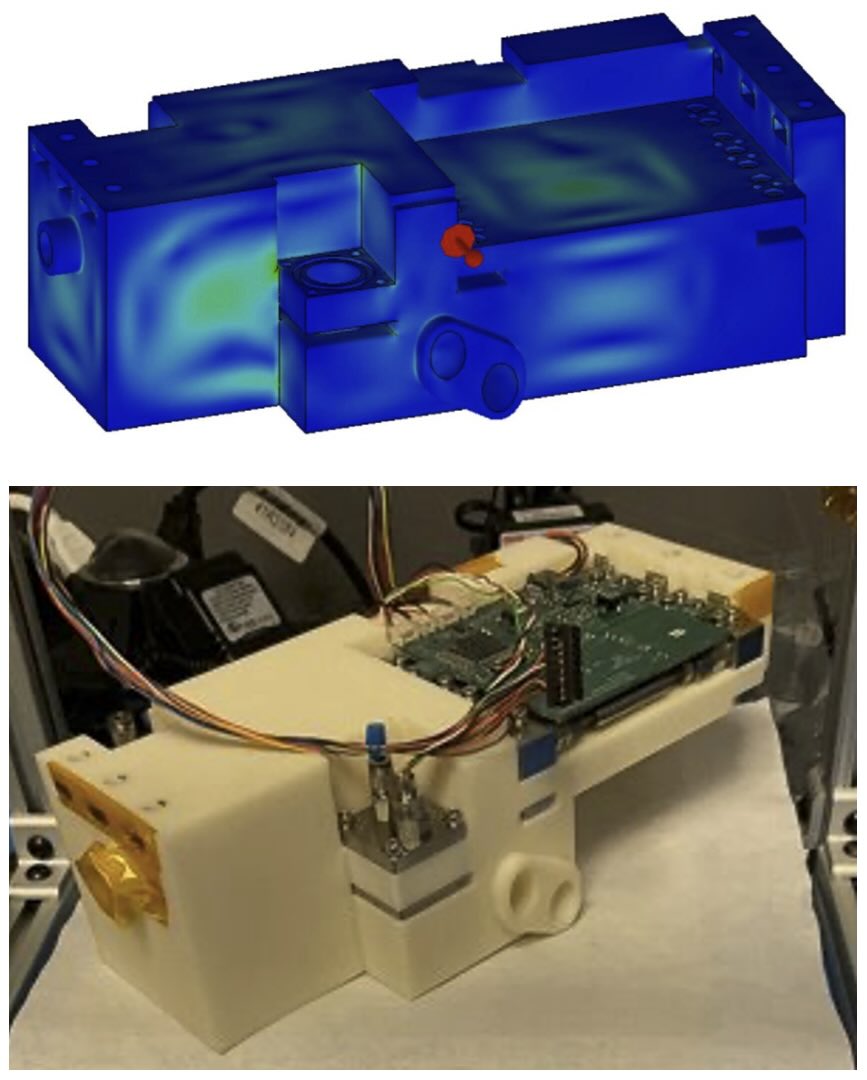}
\vspace{-.05in}
\caption{(top) 3D-printed cold gas thruster modules for VISORS \cite{hart2022}. (bottom) Actual pre-integration spacecraft module (Glenn Lightsey, private communication)}\label{fig_coldgas}.
 \vspace{-.1in}
\end{wrapfigure} 
The Georgia Tech team led by Glenn Lightsey has extensive experience using cold-gas thrusters for CubeSat control, currently as propulsion lead for the VISORS mission designed for cm-level relative positioning control.  STARI will rely on the the propulsion flight heritage from the Georgia Tech group.     Figure~\ref{fig_coldgas} shows a finite element analysis of a unit designed for VISORS as well as a pre-flight unit already constructed and undergoing final testing.


\noindent{\bf Inter-spacecraft (Cross-link) Communications}:  STARI will require high-speed, low-latency communication links between STARI-1 and STARI-2 to allow exchange of dynamic position, tip-tilt, and other telemetry, necessary for our advanced GNC algorithms. The Michigan Aerospace group led James Cutler has extensive history in developing UHF and S-Band communication links for CubeSats \cite{cutler2011,cutler2014}.  While we have endeavored to use heritage from the VISORS mission when possible, the 1 Mbps VISORS bandwidth is larger than necessary for STARI and we will develop our own system from COTS radios with less stringent requirements (100 kbps).  We motivate our bandwidth requirements assuming the VISORS GPS-based guidance systems requires $<$1~kBps (Simone D'Amico, private communication) and that each STARI spacecraft will measure 4 LED positions, 2 star positions, 1 reference laser position, tip-tilt mirror angles, along with multiple gyroscope and accelerator streams at approximately 100 Hz.  At minimum, approximately 25x 16-bit values should be exchanged at 100Hz, summing to a bandwidth of $\sim$40 kbps, easily achievable with a 10MHz cross-link bandwidth over range of 100meters. At these rates, we may not even require a special hardware solution but our primary UHF TTC radios can provide cross-link capabilities as well.  During the design process, if dedicated radios are identified, we have software defined-agile radios that can provide dedicated, higher rate communication channels. 
Our design work will identify suitable frequencies to operate on orbit between 400 MHz and 6 GHz.

\noindent{\bf Uplink and Downlink Communications:}  
We expect STARI licensing to be similar to our past CubeSat missions, experimental licenses in the amateur UHF band.  In the past, we have coordinated with the IARU for a CubeSat frequency (437.485MHz) and then applied for an experimental license in the UHF amateur radio band from the FCC.  Our satellite operators are also all licensed amateur radio operators.  We have also received experimental licenses in the 3 GHz band for downlink of high speed data.    RF spectrum licensing is an ever-changing process and we design our systems to be as flexible as possible to meet these requirements.  Our primary TTC ground station is located on the campus of the University of Michigan.

\noindent{\bf Operations:}
Our team (UM/Stanford/Georgia Tech/RPI/JPL/MIT) collectively has experience operating over a dozen CubeSat missions and will manage operations during the planned 6-month flight mission. Specifically, UM has full operations centers for both CubeSats and science on CYGNSS and SunRISE, which can be leveraged for full end-to-end CONOPS.  

\noindent{\bf Orbital Considerations: }
We will use a low-earth orbit (LEO) beyond 500km to reduce drag to acceptable levels for our propulsion system to maintain formation stability. We anticipate using similar orbits to the VISORS mission, customized for a slightly different formation geometry. VISORS maintains alignment of the spacecraft along the line-of-sight to their target (the Sun), while STARI will have the baseline separation perpendicular to the line-of-sight.  Since maintaining and an interferometry-friendly formation geometry in LEO is expensive for propellant for arbitrary orientations, we will use work from previous authors to find efficient configurations.    There is a rich history of academic work exploring this topic, going as far back as 1985 \cite{stachnik1985}, an application to TanDEM/TerraSAR-X \cite{damico2006}, and with a recent resurgence of interest \cite{hansen2020, yahya2022,ito2023}.    Our team has extensive experience planning passively-safe orbital trajectories with VISORS and Starling.   Note that a future optical space interferometer mission will likely be sited in GEO, as is already planned for the SunRISE radio interferometry mission, or for L2 which is the preferred location for the LIFE mid-infrared nulling interferometry mission.

\section{Risk Matrix and Mitigation}

\begin{wrapfigure}[14]{r}{4.5in}
\centering
\vspace{-.3in}
\includegraphics[width=4.4in]{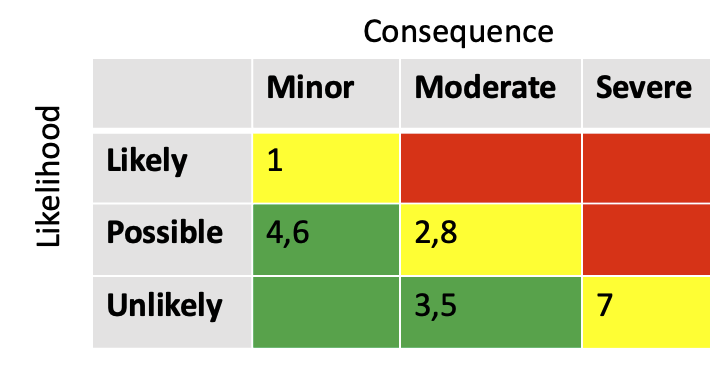}
\vspace{-.1in}
\caption{Risk Matrix (see Table\,\ref{table_risks} for itemized list)
\label{fig_riskmatrix}
}
\end{wrapfigure} 
We have identified the highest risks to the STARI Mission and presented them in risk matrix form (see Figure~\ref{fig_riskmatrix} and Table~\ref{table_risks}).  These risks are targets of the trade studies planned for formal Phase A/B studies if this project goes forward.  We note that the symmetric design of the two-spacecraft acts to both decrease the overall costs of engineering design while also offering natural redundancy.  

\begin{table}[!b]
\vspace{-.15in}
\begin{center}
\caption{\label{table_risks} Primary mission risks and mitigation plans}
\begin{tabular}{|p{.15in}|p{1.8in}|p{4in}|}
\hline \hline
 & Risk  & Mitigation \\
 \hline \hline
1 & Failure of a payload actuator or sensor & Identical STARI spacecraft provides redundancy for many systems \\
\hline
2 & Attitude control is insufficient for continuous starlight beam transfer & a) Carry-out trade to add active steering of reference laser b) Starlight injection in fiber will be statistically possible but continuous for only short times  \\
\hline
3 & One star tracker fails or is not precise enough & Precision gyroscopes provide sufficient attitude precision for partial mission success (see mitigation 2b) \\
\hline
4 & Thruster or reaction wheel impulses cause large jumps in sensors & a) Study early data from VISORS mission (same propulsion)
b) Partial mission success with unstable injection (see mitigation 2b) \\
\hline
 5   & Aperture too small; difficult finding guide stars & a) Carry-out trade to increase from 25mm$\rightarrow$50mm apertures
b) Partially rely on gyroscopes (see mitigation 2b) \\
\hline
6 & One spacecraft loses communication with ground & Add software to allow cross-link communication to provide redundancy for mission control \\
\hline
7 & One spacecraft is lost during deployment, detumbling & a) Pointing and tracking stability testing ok on survivor, but core mission goals lost b) Consider risk when choosing spacecraft bus vendor and ADCS \\
\hline
8 & Budget or schedule overruns & a) Phase B study will include JPL Team-Xc evaluation in year 2 to right-size project
b) Plan descope options, such as reduced redundancy or smaller spacecraft busses \\
 \hline
 
\end{tabular}
\end{center}
\end{table}

\section{Conclusions}

Here we present an overall design for a two-cubesat pathfinder formation flying mission that would pave the way for a formation-flying optical interferometer.  The STARI project is currently in pre-phase A and funding is being sought via NASA APRA and other programs.  The team welcomes any feedback from the community.

\newpage 
\acknowledgments 
 We acknowledge productive discussions with Michael Ireland and Ewan Douglas. We recognize seed funding from the Michigan Institute for Research in Astrophyics (MIRA).

\bibliography{report} 
\bibliographystyle{spiebib} 

\end{document}